\documentclass[a4paper]{jpconf}
\usepackage{graphicx}

\newcommand{\be}{\begin{equation}}
\newcommand{\ee}{\end{equation}}
\newcommand{\bra}{\langle}
\newcommand{\ket}{\rangle}
\newcommand{\bea}{\begin{eqnarray}}
\newcommand{\eea}{\end{eqnarray}}
\newcommand{\dis}{\displaystyle}

\begin{document}
\title{Bayesian inference with an adaptive proposal density for GARCH models}

\author{Tetsuya Takaishi}

\address{Hiroshima University of Economics, Hiroshima 731-0192, JAPAN}

\ead{takaishi@hiroshima-u.ac.jp}

\begin{abstract}
We perform the Bayesian inference of a GARCH model by
the Metropolis-Hastings algorithm with 
an adaptive proposal density.
The adaptive proposal density is assumed to be 
the Student's t-distribution and the distribution 
parameters are evaluated by using the data sampled
during the simulation.
We apply the method for the QGARCH model
which is one of asymmetric GARCH models 
and make empirical studies for  
for Nikkei 225, DAX and Hang indexes.
We find that autocorrelation times from our method are very small,
thus the method is very efficient 
for generating uncorrelated Monte Carlo data.
The results from the QGARCH model show that
all the three indexes show the leverage effect, i.e. the volatility is high
after negative observations.

\end{abstract}

\section{Introduction}
In empirical finance volatility of asset returns is an important value to measure risk. 
In order to forecast future volatility it is desirable to use appropriate models 
which have the properties of volatility  of asset returns.
Many empirical studies suggest that the distribution of asset returns 
is leptokurtic.
Furthermore the volatility is not constant, but changes over time.
There are periods when volatility is very high or very low. 
This property of the volatility is called {\it volatility clustering}.

The Autoregressive Conditional Heteroscedasticity (ARCH) model\cite{ARCH} and its generalization, 
the Generalized ARCH (GARCH) model\cite{GARCH}  
are designed to capture the property of the volatility clustering.
Moreover the distributions of returns from those models show fat-tailed distributions and  
they are suggested to be Student's t-(Tsallis) distributions\cite{DIST,DIST2}.  
There are many extensions of the GARCH model to include additional properties of the volatility.  
An example of the properties of the volatility is that the volatility response is high after negative news (returns),
which is known as the leverage effect, first observed by Black\cite{BLACK}.
In order to cope with this fact, some models\cite{EGARCH,GJR,Ding,Zakoian,QGARCH1,QGARCH2} 
which introduce asymmetry into the volatility response function are proposed. 
In this study among them we focus on the Quadratic GARCH(QGARCH) model\cite{QGARCH1,QGARCH2} which adds an additional
term proportional to a return to the volatility response function.

To utilize the GARCH models 
we need to infer GARCH parameters from financial time series data.
In general the Maximum Likelihood (ML) estimation is favored 
to the inference of GARCH models.
Although implementation of the ML method is straightforward,
there exist practical difficulties in estimating GARCH parameters by the ML technique.
The model parameters must be positive to ensure a positive volatility and
the stationarity condition for volatility is also required.
The ML method with such requirements is performed via a constrained optimization technique which can be cumbersome.
Forethermore the output of the ML method is often sensitive to starting values.

Another estimation technique is the Bayesian inference which does not have the difficulties seen in the ML method.
Usually the Bayesian inference is performed by Markov Chain Monte Carlo (MCMC) methods which have been common
in the recent computer development. 
Various MCMC methods for the Bayesian inference of the GARCH models
have been proposed\cite{Bauwens,Kim,Nakatsuma,Vrontos,WATANABE,ASAI,HMC,ARDIA}.
In a survey on the MCMC methods of the GARCH models\cite{ASAI}
it is shown that Acceptance-Rejection/ Metropolis-Hastings  (AR/MH) algorithm
with a multivariate Student's t-distribution
works better than the other algorithms.
The multivariate Student's t-distribution is used as a proposal density of the MH algorithm
and the parameters to specify the Student's t-distribution are determined by the Maximum Likelihood (ML) technique.
Recently an alternative method to estimate those parameters without relying on the ML technique was proposed\cite{ACS,ACS2,ACS3}.
The method is called "adaptive construction scheme",
where the parameters of the multivariate Student's t-distribution are determined by using the pre-sampled data by an MCMC method.
And the parameters are updated adaptively during the MCMC simulation.

The adaptive construction scheme was tested for GARCH and QGARCH models\cite{ACS,ACS2,ACS3} and
it is shown that the adaptive construction scheme can significantly reduce
the correlation between sampled data.
In this paper first we describe the adaptive construction scheme for the GARCH models.
Then we make empirical studies with the QGARCH model 
for three major stock indexes, Nikkei 225, DAX and Hang Seng.

\section{GARCH Model}
Let $x_t$ be an asset return observed at time $t$.
We transform $x_t$ to $y_t$ as
\be 
y_t=x_t-\bar{x},
\ee
where $\bar{x}$ is the average over $N$ observations, i.e. 
$\dis\bar{x} =\frac1N\sum_{i=1}^N x_i$.
In the GARCH model $y_t$ is assumed to be decomposed as
\be
y_t = \sigma_t \epsilon_t,
\ee
where $\epsilon_t$ is an identically distributed random variable with
zero mean and unit variance.
The distribution of $\epsilon_t$ is usually assumed to be a normal (Gaussian) or a leptokurtic one.
In this study we assume that the distribution is a normal one, i.e.  $\sim N(0,1)$.
In the original GARCH model the volatility $\sigma_t$ is assumed to change over time as
\be
\sigma_t^2  = \omega + \sum_{i=1}^{q}\alpha_i y_{t-i}^2
+ \sum_{i=1}^{p}\beta_i \sigma_{t-i}^2,
\label{eq:sigma}
\ee
and more specifically this model is stated as the GARCH(p,q) model. 
Since in empirical studies with the AIC analysis small numbers are often chosen 
for $p$ and $q$ we focus on the GARCH(1,1) model in this study and write it as the GARCH model.
Now eq.(\ref{eq:sigma}) is written as
\be
\sigma_t^2  = \omega + \alpha y_{t-1}^2 + \beta \sigma_{t-1}^2.
\label{eq:sigma2}
\ee

The volatility response function of eq.(\ref{eq:sigma2}) is symmetric under 
positive or negative observations $y_t$.
However some asset returns are known to show 
the leverage effect, i.e. the volatility is higher after negative observations than after positive ones.
Several extended GARCH models have been proposed to introduce asymmetry into the volatility response function.
In this study we  use the QGARCH model\cite{QGARCH1,QGARCH2} given by  
\be
\sigma_t^2  = \omega + \gamma y_{t-1}+ \alpha y_{t-1}^2 + \beta \sigma_{t-1}^2,
\ee
where the additional term, $\gamma y_{t-1}$ introduces the asymmetry into the model.
The QGARCH model includes 4 parameters ($\omega,\alpha,\beta,\gamma$) which have to be 
determined from financial data.

\section{Maximum Likelihood Estimation}
The likelihood function of the GARCH models is written as
\be
L(y|\theta)=\Pi_{i=1}^{n} \frac1{\sqrt{2\pi\sigma_t^2}}\exp\left.(-\frac{y_t^2}{2\sigma_t^2
}\right.),
\ee
where $y$ stands for the time series data of $n$ observations, $y=(y_1,....y_n)$
and $\theta$ stands for GARCH parameters, e.g. for the QGARCH model 
$\theta=(\omega,\alpha,\beta,\gamma$).

In the ML estimation the parameters are determined 
by maximizing the log likelihood $\ln L(y|\theta)$,
\be
\ln L(y|\theta) = -\frac12\sum_i^n \ln(2\pi \sigma_t^2)  
-\sum_i^n \frac{y_t^2}{2\sigma_t^2}.
\ee

\section{Bayesian Inference}

From the Bayes' theorem we obtain 
the posterior density $\pi(\theta|y)$ which is 
a probability distribution of parameters $\theta$  as
\be
\pi(\theta|y)\propto L(y|\theta) \pi(\theta),
\ee
where $\pi(\theta)$ is the prior density for $\theta$.
Since we do not know the functional form of $\pi(\theta)$ 
we make an assumption for it.
Here we assume that $\pi(\theta)$ is constant. 
Using $\pi(\theta|y)$, $\theta$ is estimated by 
\be
\bra {\bf \theta} \ket = \frac1{Z}\int {\bf \theta} \pi(\theta|y) d\theta,
\label{eq:int}
\ee
where
\be
Z=\int \pi(\theta|y) d\theta.
\ee
In general eq.(\ref{eq:int}) can not be performed analytically.
Instead of performing the integral of eq.(\ref{eq:int}) we evaluate eq.(\ref{eq:int})
by the MCMC technique described in the next section.
In the MCMC evaluation the normalization constant $Z$ will be irrelevant. 

\section{Markov Chain Monte Carlo}
The Metropolis algorithm is an MCMC technique first introduced by Metropolis {\it et al.}\cite{METRO},
which is designed to estimate an integral such as eq.(\ref{eq:int}) numerically.
The Metropolis-Hastings (MH) algorithm\cite{MH} is a generalization of the original Metropolis algorithm.
Let us consider to evaluate eq.(\ref{eq:int}) by the MH algorithm.
In general the functional form of $\pi(\theta|y)$ 
may be too complicated to generate random variables according to $\pi(\theta|y)$.
In the MH algorithm we draw random variables from a proposal density which 
is simple enough to generate random variables.
The basic procedure of the MH algorithm is given as follows. 

(1) First we set an initial value $\theta_0$ and $i=1$.

(2) Then we generate a new value $\theta_i$ from a certain proposal density $g(\theta_i|\theta_{i-1})$.

(3) We accept the candidate $\theta_i$ with a probability of $P_{MH}(\theta_{i-1},\theta_i)$
where
\be
P_{MH}(\theta_{i-1},\theta_i) = 
\min\left[1,\frac{P(\theta_i)}{P(\theta_{i-1})}\frac{g(\theta_{i-1}|\theta_i)}{g(\theta_{i}|\theta_{i-1})}\right].
\label{eq:MH}
\ee
When $\theta_i$ is rejected we keep $\theta_{i-1}$, i.e. $\theta_i=\theta_{i-1}$.

(4) Go back to (2) with an increment of $i=i+1$.

When the proposal density does not depend on the previous value,
i.e. $g(\theta_i|\theta_{i-1} ) =g(\theta_i)$ we obtain
\be
P_{MH}(\theta_{i-1},\theta_i) = \min\left[1,\frac{P(\theta_i)}{P(\theta_{i-1})}\frac{g(\theta_{i-1})}{g(\theta_{i})}\right].
\label{eq:MH2}
\ee

For a symmetric proposal density $g(\theta_i|\theta_{i-1} )=g(\theta_{i-1}|\theta_i)$, eq.(\ref{eq:MH}) reduces to 
the Metropolis algorithm and the Metropolis  accept probability is given by
\be
P_{Metro}(\theta_{i-1},\theta_i)= \min\left[1,\frac{P(\theta_i)}{P(\theta_{i-1})}\right].
\ee

Eq.(\ref{eq:int}) is evaluated as an average over the data sampled by the MCMC algorithm.  
\be
\bra {\bf \theta} \ket = \lim_{N \rightarrow \infty} \frac1k\sum_{i=1}^N \theta^{(i)},
\ee
where
$N$ is the number of the sampled data.
For $N$ independent data the  statistical error is proportional to $\frac1{\sqrt{N}}$.
However the data sampled by MCMC methods are not independent.  
For $N$ correlated data,  the  statistical error is proportional to $\sqrt{\frac{2\tau}{N}}$
where $\tau$ is the autocorrelation time between the sampled data. 
Thus in order to have a small statistical error without increasing the number of sampled data,  
it is important to take an MCMC method which generates uncorrelated data.

\section{Adaptive Construction Scheme}

The efficiency of the MH algorithm depends on how we choose the proposal density.
By choosing an adequate proposal density for the MH algorithm
one can reduce the correlation between the sampled data.
The posterior density of GARCH parameters often resembles to a Gaussian-like shape.
Thus one may choose a density similar to a Gaussian distribution as the proposal density.
Such attempts have been done by Mitsui, Watanabe\cite{WATANABE} and Asai\cite{ASAI}.
They used a multivariate Student's t-distribution in order to cover the tails of the posterior density and
determined the parameters to specify the distribution by
using the ML technique.
Here we also use a multivariate Student's t-distribution
but determine the parameters through MCMC simulations.

The ($p$-dimensional) multivariate Student's t-distribution is given by
\bea
g(\theta)& = & \frac{\Gamma((\nu+p)/2)/\Gamma(\nu/2)}{\det \Sigma^{1/2} (\nu\pi)^{p/2}} \nonumber \\
         &   & \times \left[1+\frac{(\theta-M)^t \Sigma^{-1}(\theta-M)}{\nu}\right]^{-(\nu+p)/2},
\label{eq:ST}
\eea
where $\theta$ and $M$ are column vectors,
\be
\theta=\left[
\begin{array}{c}
\theta_1 \\
\theta_2 \\
\vdots \\
\theta_p
\end{array}
\right],
M=\left[
\begin{array}{c}
M_1 \\
M_2 \\
\vdots \\
M_p
\end{array}
\right],
\ee
and $M_i=E(\theta_i)$.
$\dis \Sigma$ is the covariance matrix defined as
\be
\frac{\nu\Sigma}{\nu-2}=E[(\theta-M)(\theta-M)^t].
\ee
$\nu$ is a parameter to tune the shape of Student's t-distribution.
When $\nu \rightarrow \infty$ the Student's t-distribution goes to a Gaussian distribution.
At small $\nu$ Student's t-distribution has a fat-tail.
Since eq.(\ref{eq:ST}) is independent of the previous value of $\theta$, eq.(\ref{eq:MH2})
is used in the MH algorithm.

To use eq.(\ref{eq:ST}) as a proposal density 
we have to know the values of $M$ and $\Sigma$.
We determine these unknown parameters $M$ and $\Sigma$ as follows.
First we make a short run by an MCMC method and sample some data.
Then we estimate $M$ and $\Sigma$ from those data. 
Second substituting the estimated $M$ and $\Sigma$ to eq.(\ref{eq:ST}) 
we perform an MH simulation with the proposal density.
After accumulating more data, we recalculate $M$ and $\Sigma$, and update $M$ and $\Sigma$ of eq.(\ref{eq:ST}).
By doing this, we adaptively change the shape of eq.(\ref{eq:ST}) to fit the posterior density more accurately.
We call eq.(\ref{eq:ST}) constructed in this way "adaptive proposal density".

The random number generation for the multivariate Student's t-distribution
can be done easily as follows.
First we decompose the symmetric covariance matrix $\Sigma$ by the Cholesky decomposition as
$\Sigma=LL^t$.
Then substituting this result to eq.(\ref{eq:ST}) we obtain
\be
g(X) \sim
\left[1+\frac{X^t X}{\nu}\right]^{-(\nu+p)/2},
\ee
where $X=L^{-1}(\theta-M)$.
The random numbers $X$ are given by $\dis X=Y\sqrt{\frac{\nu}{w}}$,
where $Y$ follows $N(0,I)$ and $w$ is taken from the chi-square distribution $\nu$ degrees of freedom $\chi^2_{\nu}$.
Finally we obtain the random number $\theta$ by $\theta=LX +M$.

\section{Empirical Studies}
\begin{figure}
\vspace{5mm}
\centering
\includegraphics[height=7.0cm]{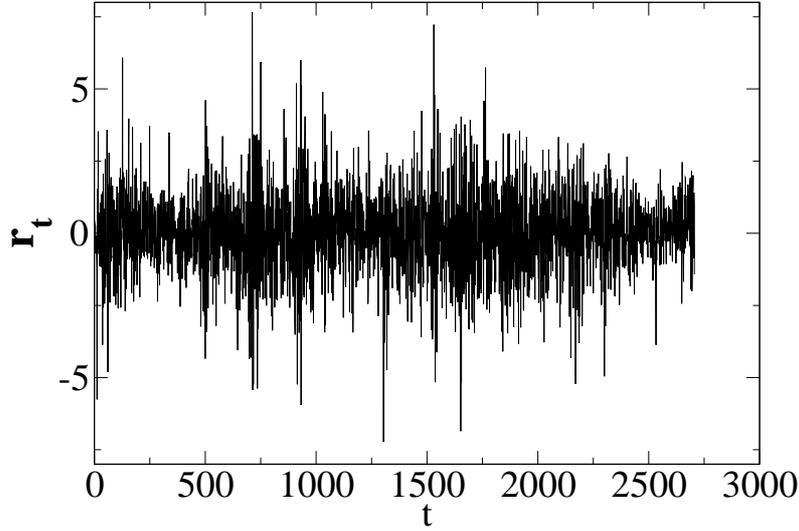}
\caption{
Time series of returns of the Nikkei 225 index.
}
\vspace{1mm}
\label{fig:nikkei}
\end{figure}

In this section we make empirical studies based on 
daily data of Nikkei 225, DAX and Hang Seng indexes.
The sampling period is  4(2)(3) January 1995  to 30 December 2005
for the Nikkei 225 (DAX)(Hang Seng) index.
The index prices $p_i$ are transformed to returns as
\be
r_i=100\ln(p_i/p_{i-1}-\bar{s}),
\label{eq:return}
\ee
where $\bar{s}$ is the average value of $\ln(p_i/p_{i-1})$.
Fig.\ref{fig:nikkei}  shows the time series of returns of the Nikkei 225 index calculated by
eq.(\ref{eq:return}) as an example.

The adaptive construction scheme is implemented as follows.
First we make a short run by an MCMC method.
Any MCMC method can be used. Here we use the Metropolis algorithm.
We discard the first 5000 data as burn-in process (thermalization).
Then we accumulate 1000 data to estimate $M$ and $\Sigma$.
The estimated $M$ and $\Sigma$ are substituted to $g(\theta)$ of eq.(\ref{eq:ST}).
The shape parameter $\nu$ is set to 10.
We re-start a run by the MH algorithm with the proposal density $g(\theta)$.
Every 1000 update we re-calculate $M$ and $\Sigma$ using  all accumulated data
and update $g(\theta)$ for the next run.
We accumulate 100000 data for analysis.
The results analyed are summarized in Table 1.

\begin{figure}
\vspace{5mm}
\centering
\includegraphics[height=7.0cm]{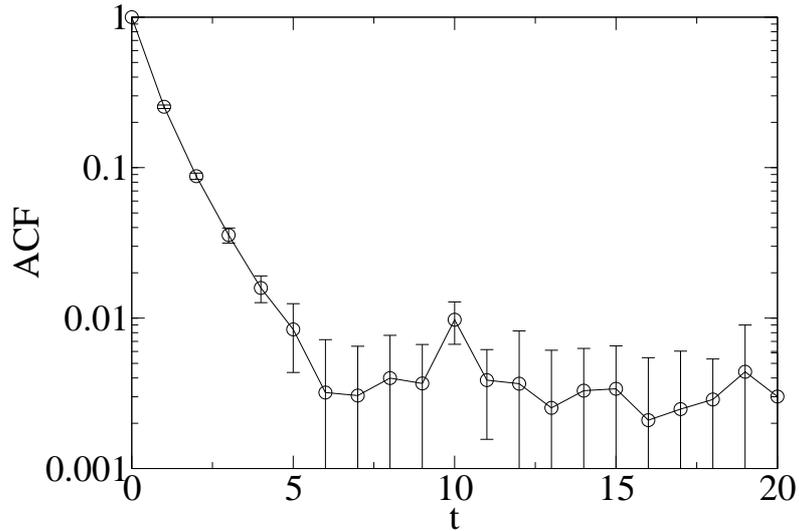}
\caption{
Autocorrelation functions of $\alpha$ sampled with  the Nikkei 225 index data.
}
\vspace{1mm}
\label{fig:ACF}
\end{figure}

We examine the efficiency of the algorithm by measuring correlations between sampled data.
Fig.\ref{fig:ACF} shows the autocorrelation function (ACF) of the sampled $\alpha$ 
with the Nikkei 225 index data.
The ACF is defined as
\be
ACF(t) = \frac{\frac1N\sum_{j=1}^N(x(j)- \bra x\ket )(x(j+t)-\bra x\ket)}{\sigma^2_x},
\ee
where $\bra x\ket$ and $\sigma^2_x$ are the average value and the variance of certain successive data $x$ respectively.
The ACF quickly decreases with $t$, which indicates that the correlation between the sampled data is very small.
We also find the similar behavior for the other parameters.

In order to analyze the autocorrelation quantitatively
we measure the integrated autocorrelation time $\tau_{int}$ given by 
\be
\tau_{int} = \frac12 +\sum_{i=1}^{\infty}ACF(i).
\ee
"$2\tau_{int}$" is also called inefficiency factor.
If no correlation exists in the data $2\tau_{int}$ takes one.
The results of $2\tau_{int}$ are summarized in Table 1.
We find that all the $2\tau_{int}$ are very small, less than 2,
which indicates that the correlation between the data sampled by this method
is small.

\begin{table}[h]
  \centering
  \caption{Results of QGARCH parameters.
   SD and SE stand for standard deviation and statistical error respectively.
Theoretically there is a relation of $SE \approx \sqrt{\frac{2\tau_{int}}{N}}SD$.
Here statistical errors are estimated by the jackknife method.}
  \label{tab:1}
  {\footnotesize
    \begin{tabular}{cllll}
      \hline
        & \multicolumn{1}{c}{$\alpha$} &
      \multicolumn{1}{c}{$\beta$} &
      \multicolumn{1}{c}{$\omega$}&
      \multicolumn{1}{c}{$\gamma$}         \\
\hline
   Nikkei 225 & 0.07872 & 0.89390 & 0.06219 & -0.12403 \\
   SD         & 0.011   & 0.013   & 0.013   &  0.021   \\
   SE         & 0.00003 & 0.00004 & 0.00005 &  0.00007 \\
   $2\tau_{int}$    & $2.0 \pm 0.1$    & $2.0 \pm 0.1$ & $2.0\pm 0.2$  & $1.8 \pm 0.1$  \\
\hline
   DAX        & 0.09198 & 0.89564 & 0.03004 & -0.08483 \\
   SD         & 0.011   & 0.011   & 0.0064  &  0.015  \\
   SE         & 0.00004 & 0.00005 & 0.00004 &  0.00006 \\
   $2\tau_{int}$    & $1.77\pm 0.06$  & $1.78\pm 0.05$  & $1.80\pm 0.07$ & $1.61 \pm 0.06$ \\
\hline
   Hang Seng  & 0.07638 & 0.91168 & 0.03202 & -0.08678 \\
   SD         & 0.009   & 0.0098  & 0.007   &  0.007  \\
   SE         & 0.00005 & 0.00005 & 0.00003 &  0.00007 \\
   $2\tau_{int}$    & $1.80\pm 0.06$  & $1.78\pm 0.06$  & $1.79\pm 0.06$ & $1.75 \pm 0.06$ \\
\hline
\hline
    \end{tabular}
  }
\end{table}

\begin{figure}
\vspace{9mm}
\centering
\includegraphics[height=7.0cm]{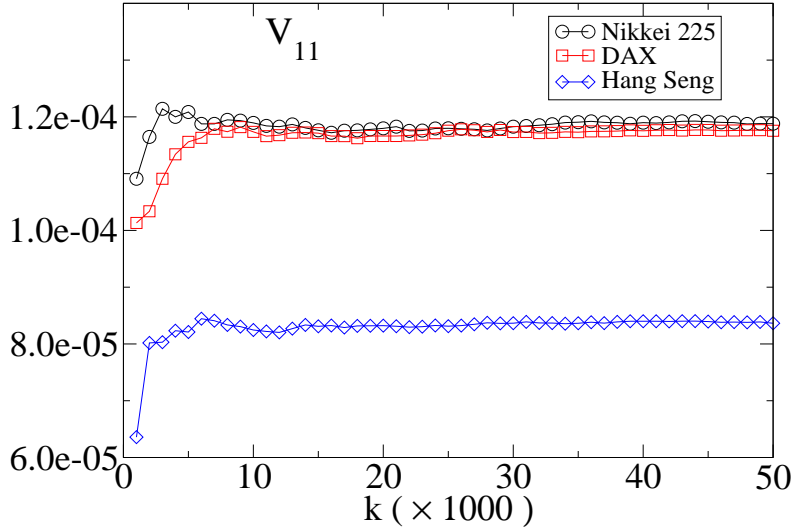}
\caption{
The diagonal element $V_{11}$ as a function of Monte Carlo time.
}
\vspace{1mm}
\label{fig:SIG11}
\end{figure}

\begin{figure}
\vspace{5mm}
\centering
\includegraphics[height=7.0cm]{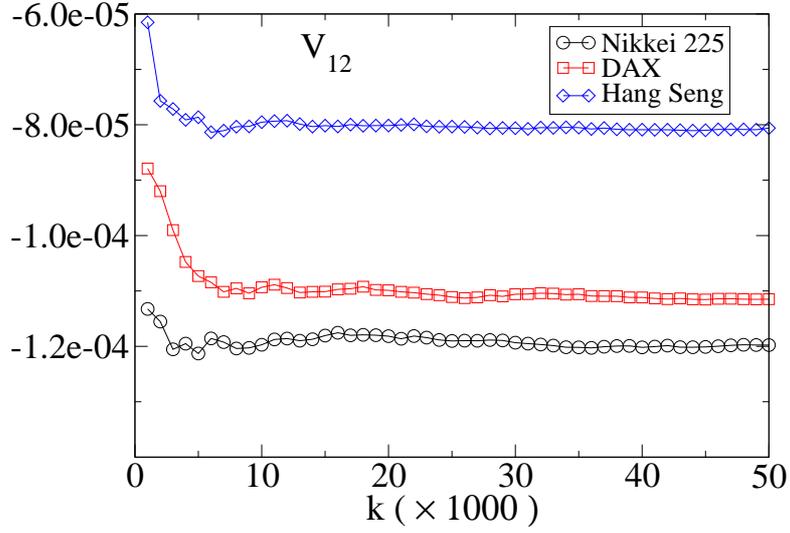}
\caption{
The off-diagonal element $V_{12}$ as a function of Monte Carlo time.
}
\vspace{1mm}
\label{fig:SIG12}
\end{figure}

Fig.\ref{fig:SIG11} and \ref{fig:SIG12} show
the convergence property of the covariance matrix.
Here $V$ is a $4 \times 4$ matrix defined by $V=E[(\theta-M)(\theta-M)^t]$.
The elements $V_{11}$ and $V_{12}$ quickly converge to 
certain values. We also see the similar behavior 
for the other elements.

\begin{figure}
\vspace{8mm}
\centering
\includegraphics[height=7.0cm]{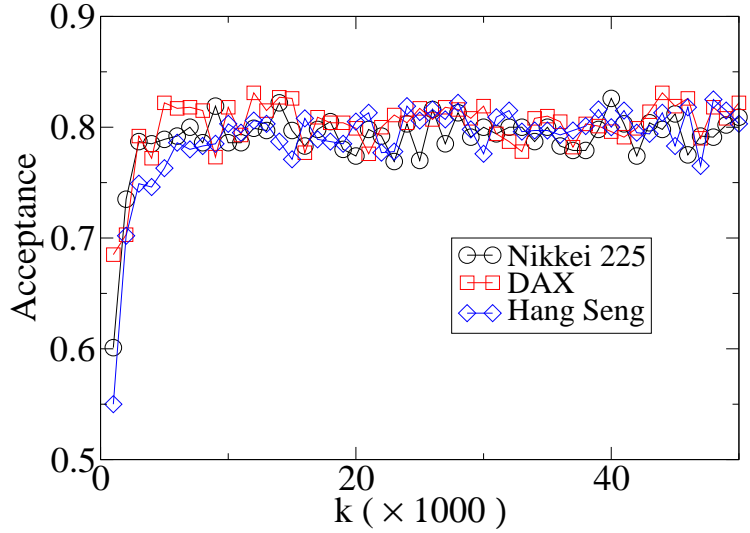}
\caption{
Acceptance at the MH algorithm with the adaptive proposal density.
}
\vspace{1mm}
\label{fig:ACC}
\end{figure}

Fig.\ref{fig:ACC}  shows the acceptance at the MH algorithm with the adaptive proposal density of eq.(\ref{eq:ST}).
Each acceptance is calculated every 1000 updates and the calculation of the acceptance is
based on the latest 1000 data.
At the first stage of the simulation
the acceptance is low. This is  because $M$ and $\Sigma$ have not yet been calculated accurately at this stage
as seen in figs.~\ref{fig:SIG11}-\ref{fig:SIG12}.
The acceptance increases quickly as the simulation is proceeded
and reaches plateaus of about $80\%$. 

\begin{figure}
\vspace{7mm}
\centering
\includegraphics[height=7.0cm]{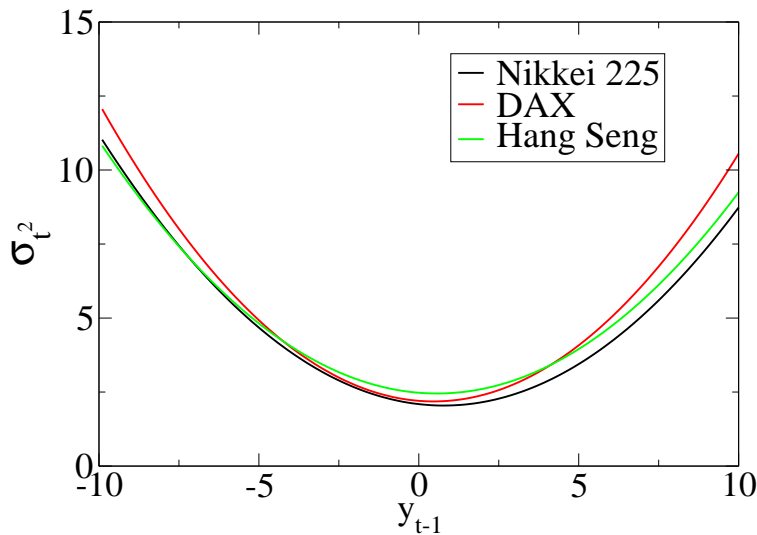}
\caption{
News impact curves of the Nikkei 225, DAX and Hang Seng indexes.
}
\vspace{1mm}
\label{fig:NIC}
\end{figure}

The QGARCH model is capable of capturing leverage effects. 
In order to see the impact of leverage effects 
Pagan and Schwert\cite{PS}, and Engle and Ng\cite{EN}
proposed the use of the so-called news impact curve.
The news impact curve is the functional relationship between
conditional variance at time $t$ and the shock at time $t-1$, $y_{t-1}$.

The news impact curve of the QGARCH model is given by
\be
\sigma_t^2 = \omega +\gamma y_{t-1} + \alpha y_{t-1}^2 + \beta\sigma^2,
\ee
where $\sigma^2$ is the unconditional variance given by
\be 
\sigma^2 =\frac{\omega}{1-\alpha-\beta}.
\ee
Fig.\ref{fig:NIC} shows the impact curves of Nikkei 225, DAX and Hang Seng indexes.
Since the values of $\gamma$ for three indexes are all negative 
the three indexes exhibit the leverage effect.
As seen in fig.\ref{fig:NIC} the news impact curve is higher for negative $y_{t-1}$ 
than for positive one. 
Since all the three news impact curves are very similar each other 
the three indexes turn out to have the similar responce to positive or negative news.

\section{Conclusions}
We have performed the Bayesian inference of the QGARCH model
by the MCMC algorithm.
The MCMC algorithm was implemented by
the MH method with the adaptive proposal density.
The adaptive proposal density is assumed to be 
the Student's t-distribution and the distribution parameters 
are determined by the data sampled by the MCMC simulation. 
The distribution parameters are updated during the MCMC simulation
adaptively to match the posterior density of the model parameters. 

We have applied our method for  
Nikkei 225, DAX and Hang Seng indexes.  
We find that the autocorrelation times between the sampled data 
are very small, typically $2\tau_{int}$ is less than 2.
Thus our method is very efficient 
for generating uncorrelated Monte Carlo data.

The QGARCH model is designed to capture the leverage effect.
We find that all the three indexes show the leverage effect,
i.e. the volatility is higher after negative observations than 
after positive ones.

The acceptance of the MH algorithm quickly reaches a plateau of about 80\% 
already at the beginning of the simulation.
This means that the distribution parameters are evaluated accurately already at 
the beginning of the simulation, as shown in figs.\ref{fig:SIG11}-\ref{fig:SIG12}.
This observation suggests that 
in practice one may stop the update of the parameters at some stage of the simulation.

\section*{Acknowledgments}
The numerical calculations were carried out on SGI Altix3700 at the Institute of Statistical Mathematics
and on NEC SX8 at the Yukawa Institute for Theoretical Physics in Kyoto University.
This study was carried out under the ISM Cooperative Use Registration (2009-ISM-CUR-0005).

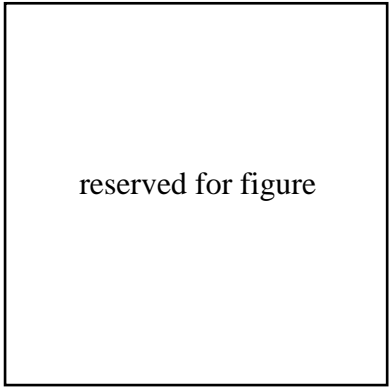
\end{document}